\documentclass{aa}             
\usepackage{graphicx}
\usepackage[T1]{fontenc}
\usepackage{amsmath}
\usepackage{graphicx}
\usepackage{txfonts}
\usepackage{natbib} 
\bibpunct{(}{)}{;}{a}{}{,} 

\begin{document} 

\title{Jupiter radio emission induced by Ganymede and \\
consequences for the radio detection of exoplanets}

\author{P. Zarka\inst{1,2}, M. S. Marques\inst{1,3}, C. Louis\inst{1}, V. B. Ryabov\inst{4}, L. Lamy\inst{1,2}, E. Echer\inst{5} and B. Cecconi\inst{1,2}}

\institute{
LESIA, Observatoire de Paris, CNRS, PSL, SU/UPD, Meudon, France
\and
USN, Observatoire de Paris, CNRS, PSL, UO, Nan\c{c}ay, France
\and
DGEF, Federal University of Rio Grande do Norte, Natal, Brazil
\and
Complex and Intelligent Systems Department, Future Univ. Hakodate, Japan
\and
INPE, Sao Jose dos Campos, Brazil
}
\date{Received <date> / Accepted <date>}

\abstract{By analysing a database of 26 years of observations of Jupiter with the Nan\c{c}ay Decameter Array, we unambiguously identify the radio emissions caused by the Ganymede-Jupiter interaction. We study the energetics of these emissions via the distributions of their intensities, duration, and power, and compare them to the energetics of the Io-Jupiter radio emissions. This allows us to demonstrate that the average emitted radio power is proportional to the Poynting flux from the rotating Jupiter's magnetosphere intercepted by the obstacle. We then generalize this result to the radio-magnetic scaling law that appears to apply to all plasma interactions between a magnetized flow and an obstacle, magnetized or not. Extrapolating this scaling law to the parameter range corresponding to hot Jupiters, we predict large radio powers emitted by these objects, that should result in detectable radio flux with new-generation radiotelescopes. Comparing the distributions of the durations of Ganymede-Jupiter and Io-Jupiter emission events also suggests that while the latter results from quasi-permanent Alfv\'en wave excitation by Io, the former likely results from sporadic reconnection between
magnetic fields Ganymede and Jupiter, controlled by Jupiter's magnetic field geometry and modulated by its rotation.}

\titlerunning{Ganymede-induced radio emission and consequences for exoplanets}
\authorrunning{Zarka et al.}
\maketitle



\section{Introduction} \label{Introduction}

Nearly 4000 exoplanets have been discovered in the past two decades (http://exoplanet.eu), but little is known yet on their interior and their rotation. It has been demonstrated that detection of their non-thermal magnetospheric radio emission will provide unique information on their magnetic field (and thus their internal structure), their rotation (directly testing spin-orbit synchronization), their orbit inclination, and the presence of satellites \citep{Hess2011,Zarka2015,Lazio2017}. Solar system exploration has revealed that the magnetospheres of Mercury, Earth, Jupiter, Saturn, Uranus, and Neptune, although resulting from the same basic plasma physics processes, show a remarkable diversity of structure and dynamics \citep{Bagenal2013}. The cyclotron maser instability (CMI) has been identified as the ubiquitous mechanism that produces the dominant high-latitude low-frequency radio emissions from these magnetospheres \citep{Zarka1998}. It is therefore expected that the detection of CMI emissions from star-exoplanet systems will shed light on their plasma interactions and open a new field of comparative exo-magnetospheric physics. Moreover, the existence of a substantial planetary magnetic field seems to favour the planet's capability to host life as it protects the atmosphere against bombardment by cosmic rays, stellar flares, and coronal mass ejections, and limits atmospheric escape \citep{Griessmeier2004,Griessmeier2005}.

In our solar system, the most intense radio emission is the decameter-wave radiation emitted by Jupiter's magnetosphere. However, although it is often as bright as solar radio bursts at frequencies below 40 MHz, it is at least $10^{3-4}$ times too weak to be detectable against the statistical fluctuations of the galactic radio background at a distance of several parsecs, even with the largest existing low-frequency radiotelescopes like UTR-2 and LOFAR \citep{Zarka2007}. The central questions conditioning radio searches for exoplanets are as follows. (1) How much can radio emissions be stronger than Jupiter's planetary -- or planet-induced -- emission? (2) How does one select the best targets to observe?

In spite of our good understanding of the CMI, there is no simple answer from first principles because the wave growth depends on the detailed distribution of keV electrons in the source and the final radio intensity depends on the source structure and size as well as on the interplay of wave convection with various saturation processes (quasi-linear diffusion, non-linear trapping). Empirical scaling laws were therefore derived for answering the above questions, in which the primary engine of the radio emission is the kinetic or magnetic power input from the solar wind to planetary magnetospheres. Average radio powers emitted by planetary magnetospheres were indeed found to be proportional to both the bulk kinetic energy flux and the magnetic energy (or Poynting) flux from the solar intercepted by the magnetosphere \citep{Zarka2001,Zarka2007}. This double correlation comes from the fact that the kinetic and magnetic energy fluxes carried by the solar wind remain in a constant ratio ($\sim$170) between the Earth and Neptune. Determining which one of the two is the real physical driver of the radio emissions is crucial for selecting observation targets. If it is the kinetic energy flux, we should select exoplanets orbiting massive stars with a large mass-loss rate \citep[e.g.][]{Ogorman2018}. If it is the magnetic energy flux, we should aim at exoplanets orbiting strongly magnetized stars \citep[e.g.][]{Folsom2016,Folsom2018}. In both cases, close-in exoplanets (hot Jupiters) should be interesting targets because both energy fluxes increase with decreasing distance to the star.

In order to determine which scaling law applies, the paradigm of solar wind--planet interaction was generalized to the interaction between a magnetized flow and a conductive obstacle (magnetized or not), leading to dissipation of the flow's power (kinetic and magnetic) on the obstacle, a fraction of which goes into electron acceleration and precipitation generating radio emissions \citep{Zarka2017b}. This paradigm can be applied to satellite-Jupiter interactions. Only the Io-Jupiter (hereafter I--J) radio emission is quantitatively documented so far, and it seems compatible with the radio-magnetic scaling law only, rough estimates only being available for the other Jovian moons \citep{Zarka2001,Zarka2007}. Unlike bodies embedded in the solar wind, interaction of Jupiter's magnetosphere with the Jovian moons is dominated by the flow of magnetic energy. This flow proceeds via Alfv\'en wave excitation at Io \citep{Saur2004}, and magnetic reconnection at Ganymede which possesses an intrinsic magnetic field \citep{Kivelson2004}.

Here we analyze the database of 26 years of radio observations of Jupiter with the Nan\c{c}ay Decameter Array, built by \citet{Marques2017}. In Sect. 2, we detail the unambiguous detection of Ganymede-Jupiter (hereafter G--J) decameter emission from this database. Over 350 G--J emission events are detected, that constitute the basis for the first quantitative study of their energetics (intensity, duration and power) (Sect. 3), that is then compared to the energetics of I--J emissions. In  Sect. 4, we combine the results obtained for I--J and G--J emissions to the scaling laws relating radio powers to incident kinetic or magnetic energy fluxes, and we show that the radio-magnetic scaling law provides a general frame for all radio emissions resulting from a flow--obstacle interaction. Then, we extrapolate this scaling law to the hot Jupiters regime and predict radio powers -- and hence flux densities -- $10^{3-7}$ times stronger than Jupiter's. Finally, in Sect. 5 we discuss emission detectability, relevance and limitations of the radio-magnetic scaling law, and further consequences of our study on the timescales of magnetic reconnection between Ganymede and Jupiter.

\begin{figure*}[ht!]
\centering
\resizebox{.9\hsize}{!}{\includegraphics{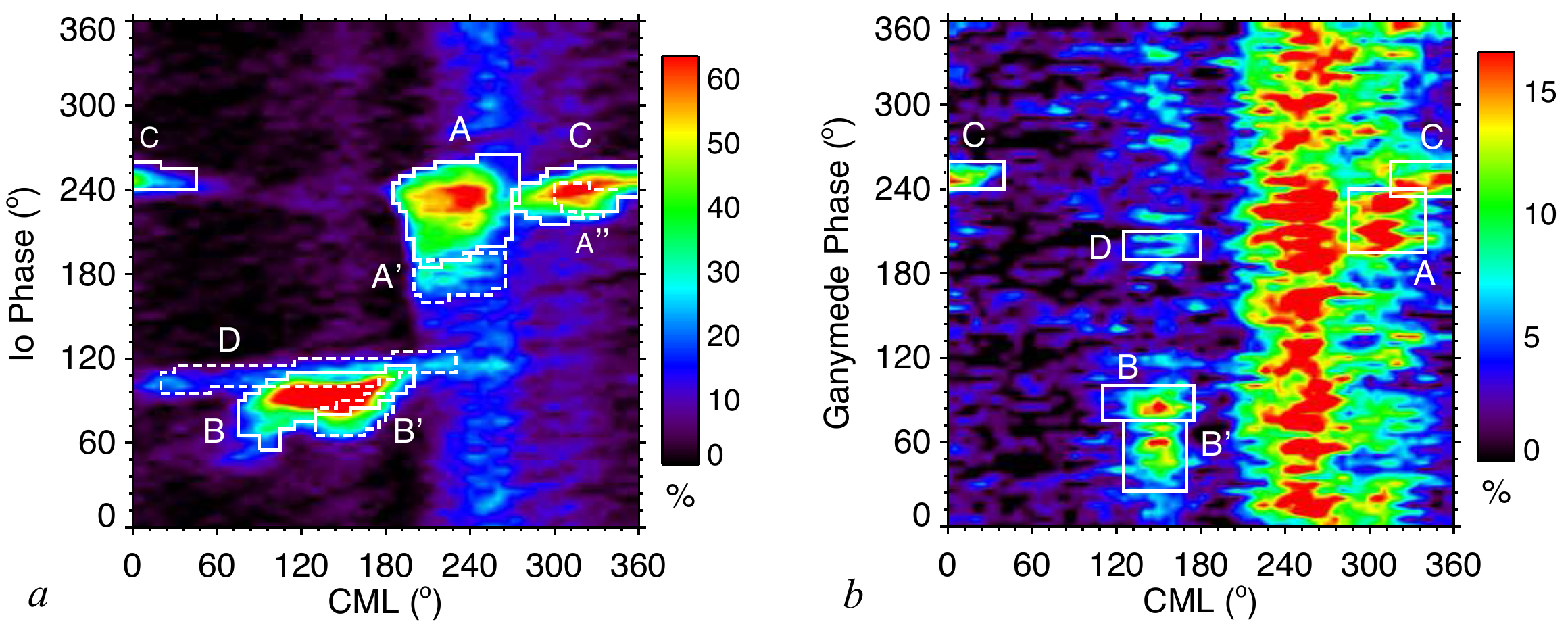}}
\caption{Occurrence probabilities of Jovian radio emissions detected over 26 years (1990-2015) with the Nan\c{c}ay Decameter Array, displayed as 2D histograms as a function of planetary rotation (CML = Central Meridian Longitude = sub-observer's longitude) and of the orbital phase of the considered satellite, in $5^\circ\times5^\circ$ bins (interpolation at $1^\circ$ resolution was applied to smooth the display). $(a)$ Occurrence probability of all emissions vs. CML and Io's orbital phase. The regions of high occurrence within letter-labelled white boxes correspond to Io-Jupiter emissions (usually named Io-A, Io-B \dots), whereas vertical bands of emission covering restricted CML ranges at all Io phases correspond to non-Io emissions (auroral or induced by other satellites). Different line styles are used to better distinguish overlapping boxes. $(b)$ Occurrence probability of non-Io emissions vs. CML and Ganymede's orbital phase. Ganymede-Jupiter emissions show up within new regions of enhanced occurrence (white boxes), labelled A to D in reference to the non-Io components in which they have been identified (see Fig. \ref{fig02} for details).}
\label{fig01}
\end{figure*}

\section{Statistical identification of Ganymede-Jupiter radio emission} \label{section 2}

I-J decametric radio emission was first identified in 1964 \citep{Bigg1964}. Similar radio emissions induced or triggered by the other Galilean satellites have been searched for with the same technique as described in \cite{Bigg1964}. Probabilities of radio emission occurrence are built as two-dimensional (2D) histograms with bins of a few degrees of observer's longitude, that is, central meridian longitude (CML) which varies with the planetary rotation, and of orbital phase $\Phi$ of the considered satellite. One influence of the satellite on the radio emissions results in regions of enhanced occurrence probability in the CML-$\Phi$ plane and in a non-uniform distribution in orbital phase. The results for Io are displayed in Figure \ref{fig01}$a$. Circular polarization sense (right-hand or left-hand) and time-frequency shape of the emission are specific for each region \citep{Marques2017}. Regions A and B correspond to the same radio source near the northern footprint of Io's magnetic flux tube in Jupiter's ionosphere, viewed from both limbs of Jupiter \citep{Queinnec1998,Marques2017}. Regions C and D similarly correspond to the source near the southern footprint of the Io flux tube. Regions A', A", and B' are extensions of regions A and B and their origin is not yet understood. 

The non-Io emissions (vertical bands of higher occurrence rate covering restricted CML ranges at all Io phases), which include auroral radio emissions as well as emissions possibly caused by satellites other than Io, also display time-frequency shapes and polarizations that can be classified in A, B, C, and D categories similar to I-J ones (i.e. northern and southern sources seen near both limbs of the planet) \citep{Marques2017}. Radio emissions induced by the other Jovian satellites have been searched among non-Io emissions. Past searches of G-J emissions in spacecraft (Voyager, Galileo, Cassini) observations \citep[and refs. therein]{Zarka2017a} provided statistical hints of their existence, in data sets spanning intervals of no longer than 2 years (the longest one having been recorded by the Galileo spacecraft). A recent re-analysis of Voyager and Cassini observations provided convincing event-by-event identifications of Ganymede- and Europa-induced radio arcs in the time-frequency plane, but without studying the intensity of these emissions \citep{Louis2017a}.

\begin{figure*}[ht!]
\centering
\resizebox{.8\hsize}{!}{\includegraphics{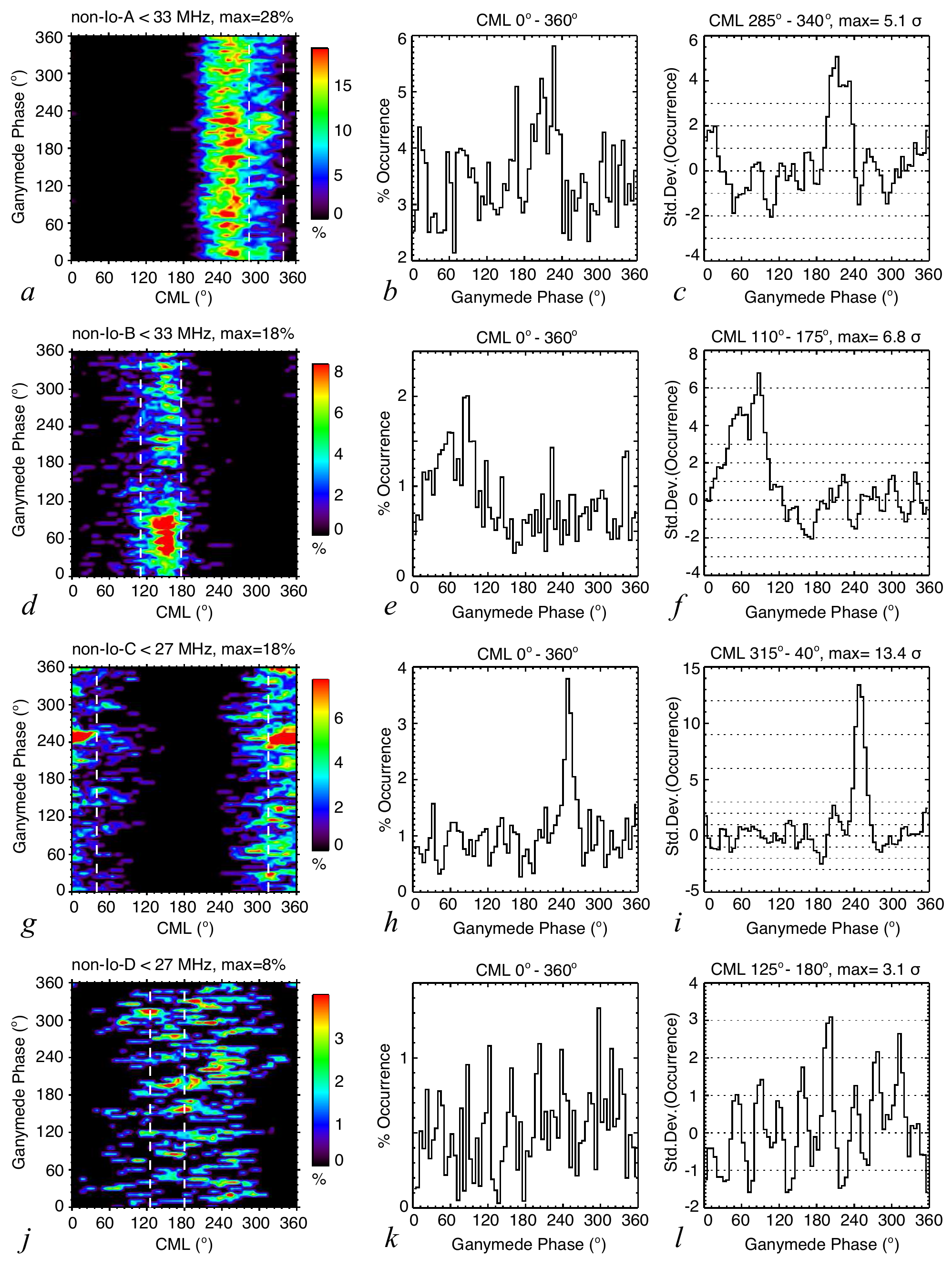}}
\caption{Occurrence probability of each component of non-Io decameter radio emissions vs. CML and Ganymede's phase $\Phi_{Ga}$. Data are non-Io emission events detected in Nan\c{c}ay between 1990 and 2015. Each component (non-Io-A, -B, -C, and -D) is searched separately for Ganymede-induced emissions. One row of plots is displayed per component. In each row, the leftmost panel shows the occurrence probability in the CML$-\Phi_{Ga}$ plane, within $5^\circ\times5^\circ$ bins (interpolation at $1^\circ$ resolution is then applied to smooth the display). The restriction in maximum frequency is explained in the text. The central panel shows the occurrence probability vs. $\Phi_{Ga}$ only (in $5^\circ$ bins), obtained by integration over all CML. The rightmost panel shows the occurrence probability vs. $\Phi_{Ga}$ obtained by integration over the CML interval delimited by the white dashed lines on the leftmost panel, and indicated above the rightmost panel. The curve in the latter panel is smoothed over three bins and ordinates are in standard deviations ($\sigma$) above the mean ($m$) ($m$ and $\sigma$ are computed in a robust way iteratively, excluding at each iteration the points $x_i$ so that $|x_i-m|>2.5\sigma$). Dotted lines indicate -3$\sigma$ to +3$\sigma$ levels by 1$\sigma$ steps, and by 3$\sigma$ steps beyond these values. The peak value is indicated above the plot. Detection of Ganymede-D emissions (panels $j, k, l$) is marginal and therefore uncertain.}
\label{fig02}
\end{figure*}

The recent construction of a 26-year-long database\footnote{The 26-year radio database from the Nan\c{c}ay Decameter Array is available in electronic form at the CDS \citep{Marques2017b} via anonymous ftp to cdsarc.u-strasbg.fr (130.79.128.5) or via http://cdsarc.u-strasbg.fr/viz-bin/qcat?J/A+A/604/A17} from daily observations of Jupiter with the Nan\c{c}ay Decameter Array from 1990 to 2015 \citep{Marques2017} allowed us to build occurrence probability diagrams versus CML and Ganymede's orbital phase (hereafter $\Phi_{Ga}$), one order of magnitude more sensitive than in previous studies due to the much broader statistical basis. In addition, because each emission event is identified individually in the database (e.g. as non-Io-A, Io-C, etc.), the search for G-J emissions can be done separately for each non-Io emission type, which proved to be more efficient. Figure \ref{fig02} shows the CML$-\Phi_{Ga}$ occurrence probability diagrams built for each non-Io component. A first analysis showed that no emission event with frequency $>$33 MHz is detected in regions of enhanced probability in the CML$-\Phi_{Ga}$ plane for non-Io-A and non-Io-B (northern) emissions. This is due to the fact that the northern footprint of the Ganymede flux tube lies northward of the northern high-amplitude magnetic anomaly at the surface of Jupiter crossed by the northern Io flux tube footprint, meaning that lower electron cyclotron frequencies are reached compared to I-J emissions. Similarly, no emission event with a frequency $>$27 MHz is detected in regions of enhanced probability in the CML$-\Phi_{Ga}$ plane for non-Io-C and non-Io-D (southern) emissions. For each non-Io component (each row in Fig. \ref{fig02}), we show first the distribution of occurrence probability in CML$-\Phi_{Ga}$ (left), then the occurrence rate versus $\Phi_{Ga}$ obtained by integration of the 2D histogram along the CML axis (centre), and finally the occurrence versus $\Phi_{Ga}$ obtained by integration only over the CML range in which a region of enhanced probability shows up in the CML$-\Phi_{Ga}$ diagram (right). 

For non-Io-A emissions (top row, panels $a,b,c$), a high-occurrence region shows up in the ranges 285$^\circ \leq$ CML $\leq340^\circ$ and $195^\circ \leq \Phi_{Ga} \leq 240^\circ$. This is detected as a broad peak in panel $b$, and clearly stands out at $>5\sigma$ level on panel $c$ restricted to the CML range $285^\circ - 340^\circ$. The CML range $\sim200^\circ - 285^\circ$ is dominated by the very high occurrence probability of auroral non-Io-A emission. In panel $c$, the Ganymede-A peak is sharp, implying that its limits in $\Phi_{Ga}$ can be well-defined. It covers the range $195^\circ \leq \Phi_{Ga} \leq 240^\circ$ at $>3\sigma$ level. Similarly, a Ganymede-B peak is clearly detected within non-Io-B emissions in the second row (panels $d,e,f$), at $>6\sigma$ level. This peak has a two-component structure, and therefore by analogy with I-J emissions we defined Ganymede-B and Ganymede-B' regions. In the third row (panels $g,h,i$), the Ganymede-C emission stands out as a prominent peak at $>13\sigma$ level. Finally in the fourth row (panels $j,k,l$), the Ganymede-D emission is tentatively and marginally detected (at $>3\sigma$ level only, around $\Phi_{Ga}\sim200^\circ$). The rectangular boxes delimiting the detected G-J regions are reproduced on Fig. \ref{fig01}$b$. The numbers of G-J emission events contained in these boxes are listed in Table 1, for each G-J emission type, and compared to the number of non-Io and Io emissions of similar type.

\begin{figure*}[ht!]
\centering
\resizebox{.8\hsize}{!}{\includegraphics{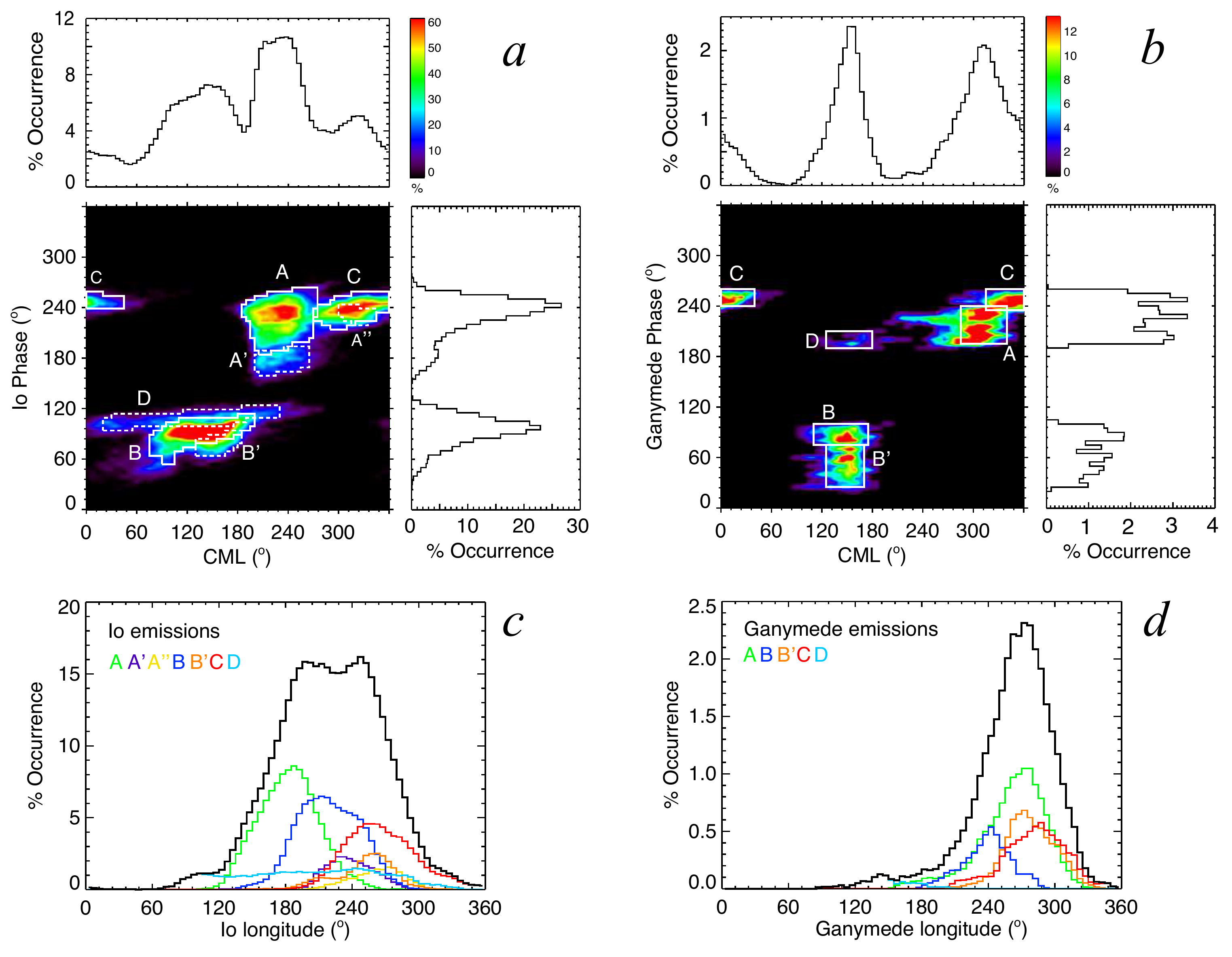}}
\caption{Occurrence probability of Io-Jupiter and Ganymede-Jupiter emissions vs. CML, phase ($\Phi$) and longitude ($\Lambda$). $(a)$ Occurrence probability of Io-Jupiter emissions only, vs. CML and $\Phi_{Io}$, with regions of enhanced occurrence identified by white boxes and labels (as in Fig. 1a). The profile integrated over all CML (resp. all $\Phi_{Io}$) is displayed on the right side (resp. top) of the colour panel. $(b)$ As in (a) for Ganymede-Jupiter emissions only, vs. CML and $\Phi_{Ga}$, with regions of enhanced occurrence identified by white boxes and labels (as in Fig. 1b). $(c$) Integrated occurrence probability of all Io-Jupiter emissions vs. Io's jovicentric longitude ($\Lambda_{Io}$ = CML+180$^\circ-\Phi_{Io}$). Components A to D from (a) are identified by colours, and their sum is the black line. $(d)$ Integrated occurrence probability of all Ganymede-Jupiter emissions vs. Ganymede's jovicentric longitude ($\Lambda_{Ga}$ = CML+180$^\circ-\Phi_{Ga}$). Components A to D from (b) are identified by colours, and their sum is the black line.}
\label{fig03}
\end{figure*}

Figures \ref{fig03}$a$ and \ref{fig03}$b$ compare the CML$-\Phi_{Io}$ distribution of I-J emissions only with the CML$-\Phi_{Ga}$ distribution of G-J emissions only. We note that bins with non-zero occurrence probability exist beyond the limits of the boxes, because when part of an emission event intersects a box, the entire event is counted as an I-J or G-J emission. The distributions show qualitative similarities, although the detailed CML and phase ranges are not identical (this was expected because these ranges result from the visibility of the radio emissions that depends on their source positions near the satellite footprints and on their three-dimensional (3D) beaming patterns). The G-J regions confirm the previous detections based on spacecraft data \citep[and references therein]{Louis2017a,Zarka2017a}, but with a much higher signal-to-noise ratio. Differences with \citet{Louis2017a} in the detailed region boundaries are attributed to the different frequency ranges of the observations studied. In Fig. \ref{fig03}$c$, the integrated occurrence probability of all I-J emissions is plotted as a function of Io's jovicentric longitude ($\Lambda_{Io}$ = CML+180$^\circ-\Phi_{Io}$). The contents of all white boxes of panel $a$ naturally merge to form a single broad peak, that corresponds to the longitude range of the physical sources of I-J radio emission in Jupiter's magnetic field. Similarly, in Fig. \ref{fig03}$d$, all G-J components merge as a single peak of occurrence probability versus Ganymede's longitude ($\Lambda_{Ga}$ = CML+180$^\circ-\Phi_{Ga}$), strongly supporting the correct identification of G-J emission events. Although marginally detected, the Ganymede-D component contributes to the left side of the curve in panel $d$, analogous to Io-D emission in panel $c$, and therefore we retained it in our analysis (it only represents a small number of G-J events, 11 out of 362, see Table 1). We have also checked that G-J emission events are quasi-uniformly distributed across the 26-year interval studied, and that their distribution shows no clustering versus $\Phi_{Io}$, which might have been due to misidentification of a small fraction of I-J events as non-Io events, whose clustering versus $\Phi_{Ga}$ would result from the 1:4 orbital resonance of Ganymede with Io.

\begin{table}
\caption{Number of emission events per type detected in Nan\c{c}ay between 1990 and 2015 \citep{Marques2017,Lamy2017}. The numbers are listed separately for types A (+A' and A"), B (+B'), C and D, for Io-Jupiter emissions (cf. Fig. \ref{fig03}$a$), non-Io emissions (cf. Fig. \ref{fig01}$b$, excluding Ganymede-Jupiter regions), and Ganymede-Jupiter emissions (cf. Fig. \ref{fig03}$b$). Each event has a duration of a few tens of minutes and counts as 1 in the histograms of Fig. \ref{fig05}.} 
\label{Table1} 
\centering 
\begin{tabular}{c c c c} 
\hline\hline 
                &  Io-Jupiter  &   Non-Io        &  Ganymede-Jupiter  \\ 
                & emissions  &   emissions   &  emissions   \\
\hline 
A (+A' +A")   & 1170  & 2012 & 122      \\
B (+B')         & 788    & 665   & 167    \\
C                   & 368    & 647   & 62        \\
D                   & 265    & 289   & 11        \\
Total             & 2591  & 3613 & 362   \\
\hline 
\end{tabular}
\end{table}

\section{Energetics (intensity, duration and power) of Io, non-Io, and Ganymede radio emissions}

\begin{figure*}[ht!]
\centering
\resizebox{.9\hsize}{!}{\includegraphics{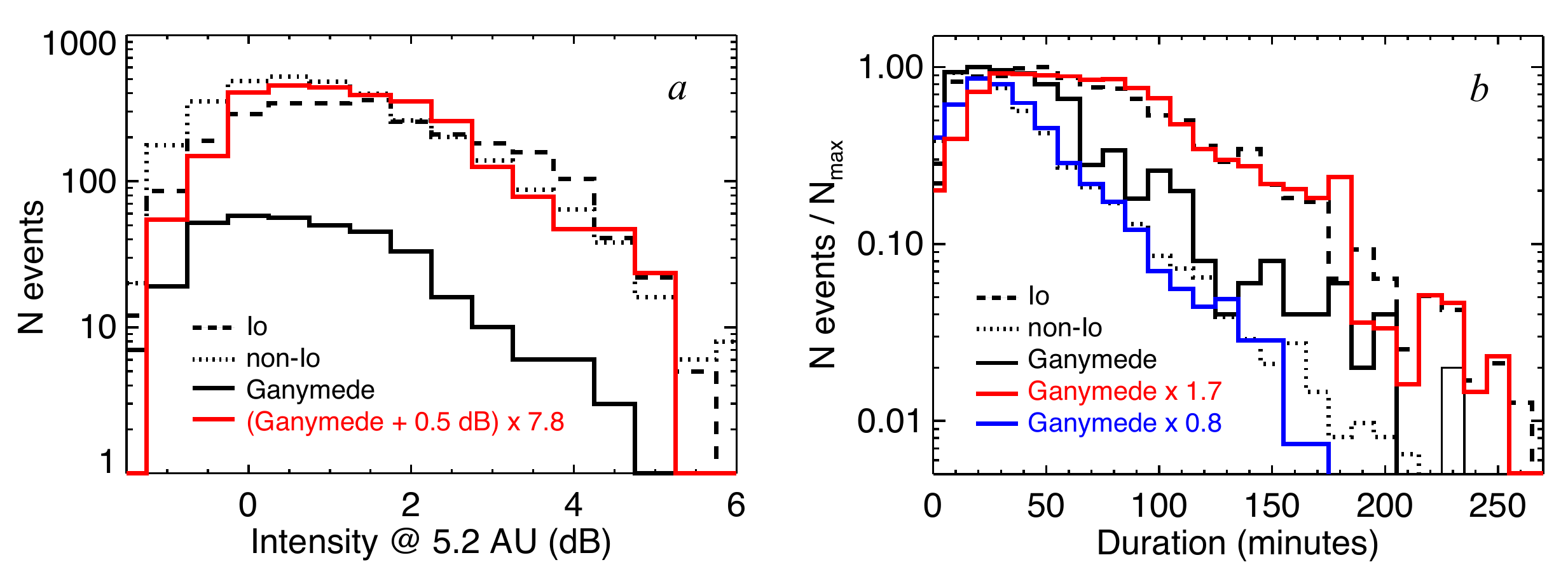}}
\caption{Intensities and durations of Ganymede-Jupiter radio emissions compared to Io-Jupiter and non-Io ones. $(a)$ Distributions of intensities (in dB above the Galactic background) of Io-Jupiter, non-Io and Ganymede-Jupiter radio emission events, normalized to an Earth-Jupiter distance of 5.2 AU (in black). A rescaled distribution of Ganymede-Jupiter radio emissions matching Io and non-Io ones is displayed in red. $(b)$ Durations of Ganymede-Jupiter radio emissions compared to Io-Jupiter and non-Io ones. Normalized distributions of durations of Io-Jupiter, non-Io, and Ganymede-Jupiter radio emission events, corrected from the variable Earth-Jupiter distance, are displayed in black. Rescaled distributions of Ganymede-Jupiter radio emissions matching Io-Jupiter and non-Io ones are displayed in colour.}
\label{fig04}
\end{figure*}

Comparison of the energetics of I-J, G-J, and non-Io emissions is based on the distributions of intensities and durations of these emissions. Both quantities vary with the observer-Jupiter distance $R$. The intensity varies in $1/R^2$, with $R$ varying between 3.95 and 6.45 AU (5.2 AU on average) over the $\sim$13-month synodic period of the Earth-Jupiter system. The intensity of each emission event can be normalized to the average distance of 5.2 AU by multiplying by $(R/5.2)^2$. Figure \ref{fig04}$a$ displays the distributions of intensities of I-J, non-Io, and G-J events, normalized to an Earth-Jupiter distance of 5.2 AU.  As noted in Fig.
8$a$ of \citet{Marques2017}, I-J and non-Io distributions look very similar. The G-J distribution contains far fewer events (see. Table 1), and is slightly shifted toward lower intensities. In order to quantitatively assess the difference between the G-J and the I-J or non-Io distributions, the distribution of G-J emission intensities was shifted by intervals of 0.5 dB, and for each shift the resulting distribution was divided bin-to-bin by the I-J or non-Io distributions. We found that for a shift of +0.5 dB, the obtained ratios are flat curves with an average value of 1/7.8, demonstrating that G-J radio emissions are $\sim$7.8 times less frequent but only 0.5 dB weaker on average than I-J and non-Io emissions.

\begin{figure*}[ht!]
\centering
\resizebox{.9\hsize}{!}{\includegraphics{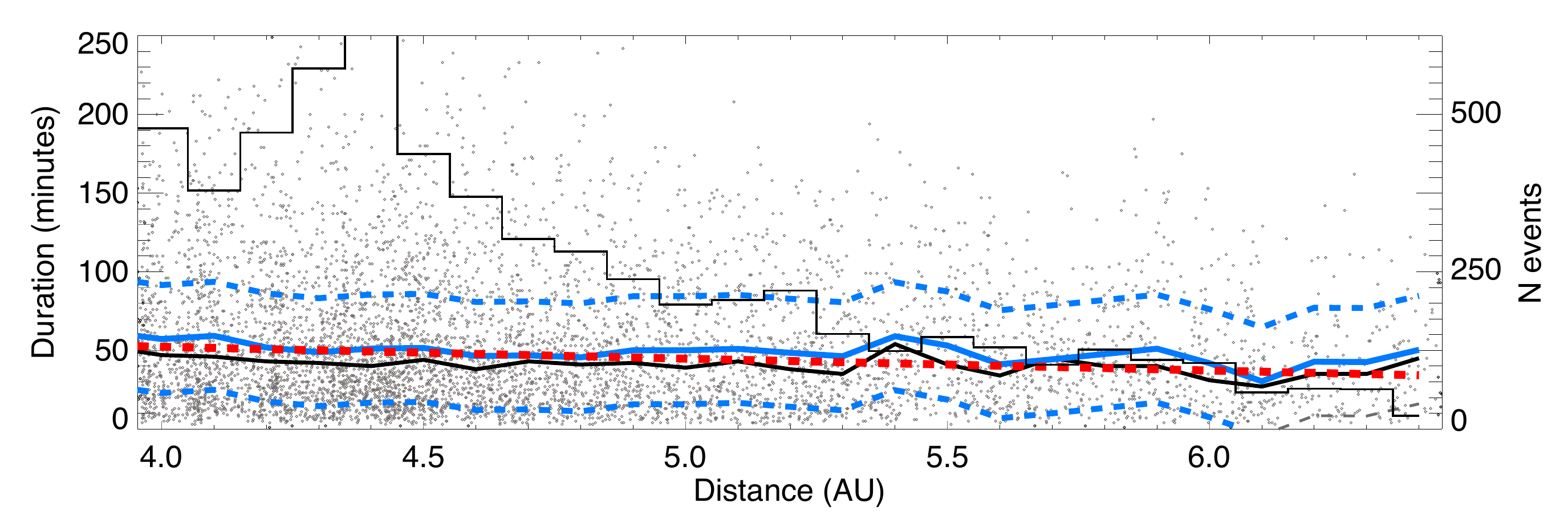}}
\caption{Emission durations vs. Jupiter-observer distance. Dots represent the duration of all emission events in our 26-year database. The histogram in black, corresponding to the scale on the right side, displays the number of dots per 0.1 AU interval. The black solid line is the median duration per 0.1 AU interval. The blue solid line is the robust mean duration per 0.1 AU interval (computed by excluding iteratively the points outside of a $m\pm3\sigma$ interval), and the blue dashed lines indicate $m\pm1\sigma$. The red dashed line is the linear fit to the median and robust mean over the interval [3.9, 5.2] AU (that contains most data points); it corresponds to a duration of 43.2 min at 5.2 AU and a slope of -7.6 min/AU.}
\label{fig05}
\end{figure*}

It is less straightforward to quantitatively compare the distributions of I-J, G-J, and non-Io emissions durations because the corresponding histograms have different shapes (widths). The event duration depends on the detection threshold above the galactic and instrumental backgrounds combined with the intensity variation of the Jovian emissions, which itself depends on the distance $R$. As a first step, we attempt to correct the measured event durations for this bias. Figure \ref{fig05} shows the distribution of the durations of all emissions in our 26-year database, from which we derive a linear approximation of the variation of the average emission duration $<D>$ as a function of $R$:
\begin{equation} \label{eq1}
<D> (min.) = 43.2 - 7.6 \times ( R - 5.2 AU )
.\end{equation}
We can use it to correct the duration of each emission event to first order by computing:
\begin{equation} \label{eq2}
D_{cor} = D +7.6 \times ( R - 5.2 AU )
,\end{equation}
with $D$ and $D_{cor}$ in minutes. We finally obtain in Figure \ref{fig04}$b$ the statistical distribution of event durations normalized to the average distance of 5.2 AU. All of the following plots and discussions refer to these corrected durations only.

\begin{figure*}[ht!]
\centering
\resizebox{.7\hsize}{!}{\includegraphics{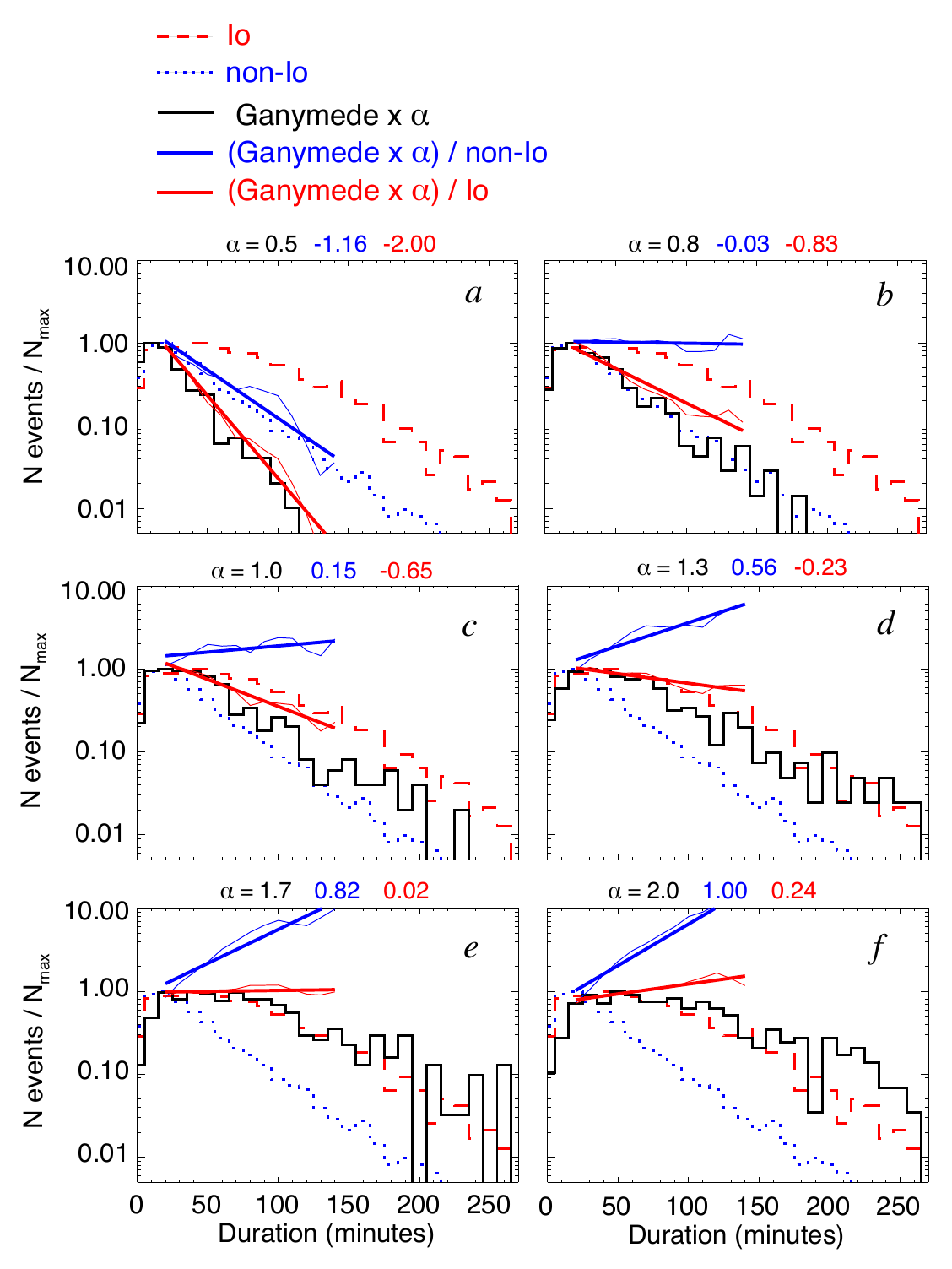}}
\caption{Determination of the ratio between typical durations of Ganymede-Jupiter, Io-Jupiter, and non-Io emissions. Histograms of rescaled Ganymede-Jupiter event durations by a factor $\alpha$ are displayed for six representative values of $\alpha$ -- indicated above each panel -- among those studied (0.5 $\leq \alpha \leq$ 2.0, by steps of 0.1). On all panels, the red dashed (resp. blue dotted) histogram is the normalized distribution of durations of Io-Jupiter (resp. non-Io) emission events, similar to the black dashed and dotted histograms of Fig. \ref{fig04}$b$. The black solid line is the normalized distribution of rescaled durations of Ganymede-Jupiter events. Thin solid red (resp. blue) lines are the bin-to-bin ratio of the rescaled Ganymede-Jupiter histogram to the Io-Jupiter (resp. non-Io) one, over the range 20-180 minutes. Boldface straight lines are the degree 1 (lin-log) fit of the thin lines of corresponding colour, whose slopes are indicated at the top of the plots.}
\label{fig06}
\end{figure*}

To compare the normalized distributions (i.e. with an histogram maximum=1) of I-J, non-Io, and G-J event durations of Fig. \ref{fig04}$b$, the question is to find which factor to apply to the G-J event durations distribution to match the I-J or non-Io ones. A simple ratio of the histograms will not provide the correct answer. We have therefore scaled G-J event durations by a factor $\alpha$ (with $0.5 \leq \alpha \leq 2.0$ in steps of 0.1). For each value of $\alpha$, the duration of every G-J event is multiplied by $\alpha$ before building the normalized histogram of G-J event durations (in black in Figure \ref{fig06}). The bin-to-bin ratio of this rescaled histogram with the histograms of I-J (dashed red) and non-Io (dotted blue) event durations is computed over the range 20-180 min, that is, between the peak of the histograms and the abscissa at which the number of events per ten-minute bin becomes $\leq$2. This ratio is plotted as the thin solid red line for the $\alpha$(G-J)/(I-J) event durations ratio, and as the thin solid blue line for the $\alpha$(G-J)/(non-Io) event durations ratio. It is fitted with straight lines (in lin-log scale) whose slopes are indicated with the same colour code at the top of the plots. When $\alpha$ varies from 0.5 to 2.0, the slopes regularly increase (the straight line fits turn counterclockwise). The value of $\alpha$ for which the red (respectively, blue) slope is closest to zero is the best estimate of the typical ratio of G-J to I-J (resp. to non-Io) event durations. We find $\alpha$=1.70 ($\pm$0.05) for the (G-J)/(I-J) ratio (panel $e$ of Fig. \ref{fig06}, where the normalized histogram of rescaled G-J event durations matches well the normalized histogram of I-J event durations) and $\alpha$=0.8 ($\pm$0.05) for the (G-J)/(non-Io) ratio (panel $b$, where the normalized histogram of rescaled G-J event durations matches well the normalized histogram of non-Io event durations).

Combining the results of Fig. \ref{fig04}$a$, which show that G-J radio emissions are $\sim$7.8 times less frequent but only 0.5 dB weaker on average than I-J and non-Io emissions, with Fig. \ref{fig04}$b$, which shows that the typical duration of G-J radio emissions is $\sim$1.7 times shorter than that of I-J emissions, allows us to compute the average power of G-J radio emissions, which is $1.7\times7.8\times10^{0.05}$  , approximately $15$ times lower than the average power of I-J radio emissions. 

\section{Scaling laws}

The kinetic and magnetic energy fluxes from Jupiter's magnetosphere intercepted by Io and Ganymede can be computed from the plasma parameters in the satellites' vicinity, as measured by the Voyager and Galileo spacecraft \citep{Kivelson2004,Zarka2007}: 
\begin{equation} \label{eq3}
P_{kin} = (\rho V^2) V \pi R_{obs}^2
,\end{equation}
and
\begin{equation} \label{eq4}
P_{mag} = (B^2/\mu_\circ) V \pi R_{obs}^2
,\end{equation}
with $\rho$ and $B$ being the (sub-)corotating jovian plasma density and magnetic field amplitude at the satellite orbit, and $V$ the flow velocity:
\begin{equation} \label{eq5}
V = V_{corot} - V_{orb} = \eta \Omega_J L R_J - (GM_J/LR_J)^{1/2}
,\end{equation}
where $\eta<1$ characterises sub-corotation, $\Omega_J$, $M_J,$ and $R_J$ characterise Jupiter's rotation, mass, and radius, and $LR_J$ is the satellite orbit radius. $\pi R_{obs}^2$ is the cross-section of the obstacle, that is, Io's ionosphere ($R_{obs} \sim 1.1 \times R_{Io}$) or Ganymede's magnetosphere ($R_{obs} \sim 2.5 \times R_{Ga}$). Using the measured parameter values from Table 21.1 of \cite{Kivelson2004}, one finally obtains the ratios $<P_{kin}(Io)/P_{kin}(Ga)>$ $\sim 5$ (with a large variability between 3 and 200 around this average value) and  $<P_{mag}(Io)/P_{mag}(Ga)>$ $\sim17$  (with a lower variability between 10 and 40 around this average value). The observed ratio between the I-J and G-J radio powers, found to be $\sim$15 above, is therefore consistent with the ratio of the magnetic energy (Poynting) fluxes intercepted by Io and Ganymede.

\begin{figure*}[ht!]
\centering
\resizebox{1.\hsize}{!}{\includegraphics{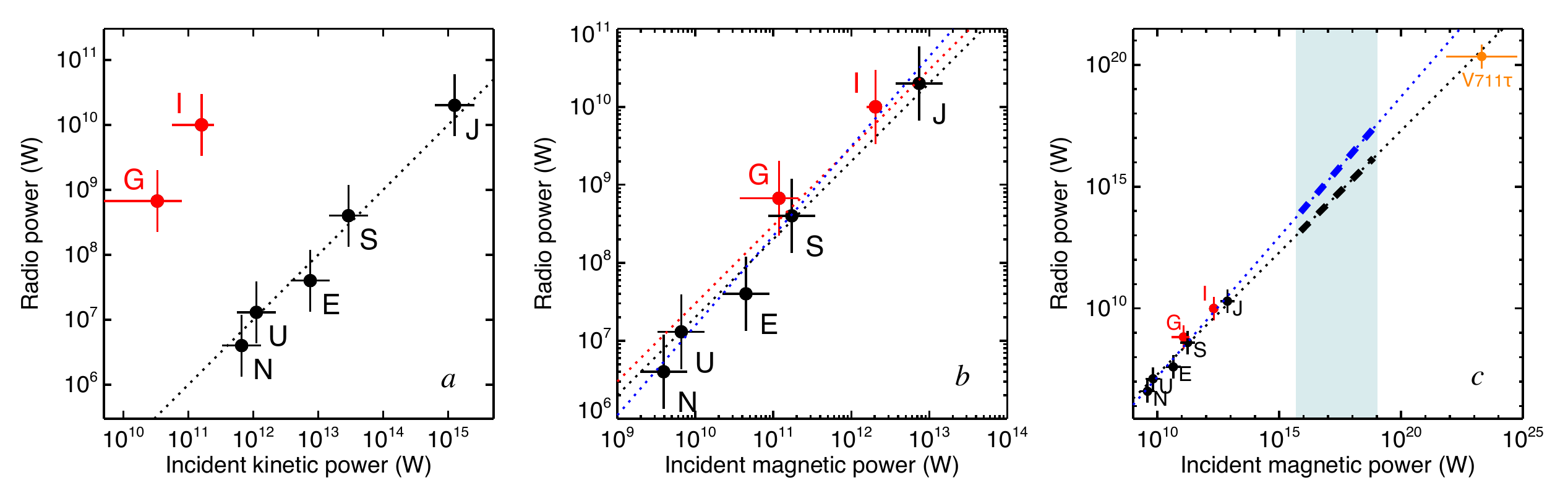}}
\caption{Radio-kinetic and radio-magnetic scaling laws and extrapolation to hot Jupiters. $(a)$ Radio-kinetic scaling law relating the overall power of the auroral radio emissions from (E)arth, (J)upiter (non-Io emissions), (S)aturn, (U)ranus and (N)eptune (integrated over their full spectrum, beaming solid angle, and time-averaged) to the bulk kinetic energy flux from the solar wind intercepted by the corresponding magnetospheric cross-sections (black dots and labels, taken from \citet{Zarka2007}). The dotted line fit has a slope of 1 and a conversion efficiency of $10^{-5}$. The red dots and labels relate the Io-induced and Ganymede-induced radio emission powers to the bulk kinetic energy flux of (sub)corotating plasma within Jupiter's magnetosphere. Error bars are estimated from the measured fluctuations of all displayed quantities \citep{Kivelson2004}. $(b)$ Radio-magnetic scaling law, relating the same radio powers as in $a$ with the incident Poynting flux from the solar wind onto planetary magnetospheres or from Jupiter's magnetosphere on Io's ionosphere and Ganymede's magnetosphere. The black dotted line, which fits the planetary magnetosphere points (black dots) well, has a slope of 1 and a conversion efficiency of $2\times10^{-3}$ (from \cite{Zarka2007}). Including the satellites (red dots), a better fit with a slope of 1 is obtained with a constant $3\times10^{-3}$ (red dotted line). The best fit with unconstrained slope has a slope of 1.15 (blue dotted line). $(c)$ Extrapolation of the radio-magnetic scaling laws from $b$ to the parameter ranges for hot Jupiters (shaded region), where the intercepted Poynting flux can reach $10^3$ to $10^6$ times that intercepted by Jupiter (taking into account increased magnetospheric compression and a solar-like magnetic field of $\sim$1 Gauss). With the law of slope 1, expected radio powers are similarly $10^{3-6}$ times stronger than Jupiter's, and one order of magnitude larger with the law of slope 1.15. Stronger stellar magnetic fields should lead to stronger radio emissions. The documented case of the magnetic binary V711$\tau$, where the involved magnetic fields are much stronger than the Sun's, is plotted as an orange point.}
\label{fig07}
\end{figure*}

Figure \ref{fig07} displays the radio-kinetic ($a$) and radio-magnetic ($b$) scaling laws derived from planetary magnetospheric auroral emissions in the solar system (black dots). For planetary magnetospheres, the incident energy fluxes come from the solar wind and are intercepted by the magnetosphere's cross-sections. Placing on these diagrams the I-J and G-J measured radio powers versus the intercepted kinetic and magnetic powers by Io and Ganymede computed above, it appears clear that the radio-magnetic scaling law provides a far more consistent paradigm for flow-obstacle interactions leading to non-thermal radio emissions. The points in Fig. \ref{fig07}$b$ are well described by a radio-magnetic scaling law with a slope of 1 and average conversion efficiency of the incident Poynting flux into radio emission power $P_{radio}/P_{mag} = 3\times10^{-3}$. The unconstrained best fit is obtained with a slope of 1.15. Satellite-Jupiter points are located slightly above the fitted law, possibly because the sub-Alfv\'enic Jovian flow is more efficiently tapped for accelerating electrons than the super-Alfv\'enic solar wind. 

We note that observations and modelling by \citet{Chane2012,Chane2015} showed that the interaction between solar wind and Earth can become temporarily sub-Alfv\'enic, with disappearance of the bow shock, apparition of Alfv\'en wings in the solar wind supported by the Earth's magnetopause -- very similar to Ganymede's Alfv\'en wings in the Jovian magnetosphere --, and reduction of auroral currents.

Figure \ref{fig07}$c$ shows the extrapolation of the two fits of Fig. \ref{fig07}$b$ to the range of Poynting flux intercepted by hot Jupiters, which would lead to radio emissions up to $10^7$ times stronger than Jupiter's, intense enough to be detectable with UTR-2, LOFAR, and SKA from exoplanetary systems at distances of several tens of parsecs \citep{Zarka2007,Zarka2015}. One possible drawback would be the existence of a saturation of the radio-magnetic scaling law, but at least one documented case exists of a magnetic binary star (V711$\tau$) where observations \citep{Budding1998,Richards2003} allowed \citet{Mottez2014} to estimate $P_{radio} \sim 7\times10^{19-20}$ W and $P_{mag} \sim 7\times10^{21} - 6\times10^{24}$ W. The corresponding point falls close to the radio-magnetic scaling law of slope 1, at powers $>10^{10}$ times larger than in the solar system, suggesting that saturation is not a critical issue.

\section{Discussion}

Based on the recent 26-year database of Jupiter observations with the Nan\c{c}ay decameter array built by \citet{Marques2017}, the prominent I-J component was revisited (Figs. \ref{fig01}$a$ and \ref{fig03}$a,c$) and the G-J component was detected unambiguously \citep[Figs. \ref{fig01}$b$ and \ref{fig03}$b,d$, and][]{Zarka2017a}. More than 350 G-J emission events were detected in the interval 1990-2015 (Table 1), making it possible to statistically characterise their duration, intensity, and, therefore, energetics (average power), and compare them to those of I-J emission events.

We have found that G-J emissions have typical intensities only $\sim$0.5 dB lower than I-J or non-Io emissions (Fig. \ref{fig04}$a$). This suggests that CMI operates at relatively uniform efficiency around Jupiter, where the various radio components are produced, whatever the origin of the accelerated electrons, leading to similar intensity distributions for all radio components. However, the temporal behaviours of G-J and I-J emissions are very different. 

G-J emissions are much (7.8 times) less frequent than I-J emissions (Fig. \ref{fig04}$a$), and have a typical duration $\sim$1.25 times longer than that of non-Io (auroral) emissions, but $\sim$1.7 times shorter than that of I-J emissions (Fig. \ref{fig04}$b$). These properties can shed light on the physics of the interaction of Ganymede and Jupiter via magnetic reconnection.
Non-Io emission events are believed to be associated with "hot spots" (localized precipitations) along Jupiter's main auroral oval \citep{Bagenal2017}. As CMI emission is strongly anisotropic (beamed in a hollow conical sheet widely open around the magnetic field within the source), Jupiter's rotation carries the beam out of the observer's view in a few tens of minutes. I-J emission being tied to Io's flux tube, the radio beam moves with Io's orbital motion, four times slower than Jupiter's rotation, which explains a duration of I-J events much longer than non-Io ones, assuming that the I-J emission is produced in a quasi-permanent way \citep{Louis2017b}. If the G-J radio emission was permanent, emission events would last even longer due to the slower orbital motion of Ganymede \citep{Louis2017a}. Figure \ref{fig04}$b$ shows that this is not the case, and that G-J emission events are likely controlled by Jupiter's rotation. We propose that the G-J interaction via reconnection is governed by a substorm-like regime of storage and sporadic release of energy controlled by Jupiter's rotation, in contrast with an I-J interaction governed by more steady excitation of Alfv\'en waves. Such waves are also likely produced in the wake of Ganymede, following the last reconnection of Jovian magnetic field lines with Ganymede's magnetosphere, but they do not seem to generate detectable emissions comparable to (and longer than) I-J ones.

Our main result is that Ganymede- and Io-induced radio powers are in the same ratio as the magnetic power input that they intercept from the magnetosphere, in spite of the different interactions of these moons with Jupiter's magnetosphere (primarily via Alfv\'en waves for Io and magnetic reconnection for Ganymede). Auroral, Io-induced, and Ganymede-induced radio emissions are all found to fit a radio-magnetic scaling law. Quantitative inclusion of G-J and I-J radio emissions strongly grounds this scaling law, the extrapolation of which allows us to predict strong -- potentially detectable -- radio emissions from hot Jupiters. 

Recent theoretical works \citep{Nichols2011,Saur2013,Nichols2016} that examined specific cases of flow-obstacle magnetic interaction generally agree with the radio-magnetic law, although their quantitative predictions for exoplanets may differ by one order of magnitude for giant planets, and up to two for Earth-like planets, over  a total range $\geq$10 orders of magnitude covered by the scaling law. 
Figures \ref{fig07}$b,c$ characterise average powers for planets orbiting a solar-type star (i.e. with a solar-like magnetic field of $\sim$1 Gauss). Stronger stellar magnetic fields should lead to stronger radio emissions. Intrinsic variability of radio emission is also superimposed on the average behaviour of Fig. \ref{fig07}$b,c$ and may lead to stronger radio bursts. Radio scintillation can temporarily further increase the received flux density by $>$1 order of magnitude.
Overall, detectable emissions levels should exist for at least a fraction of the known hot Jupiters, provided that high enough frequencies are emitted (above a few 10's MHz). Very favourable targets are hot Jupiters orbiting stars more strongly magnetized than the Sun, where radio emission can be excited by the planet interaction with the star's magnetic field in a giant analogue of the I-J or G-J systems (and for which the predicted radio power is also increased). This suggests that radio detection of exoplanets should occur soon provided that enough hot Jupiter targets are monitored, which will be the case with the deep surveys of LOFAR \citep[ongoing,][]{Shimwell2017} and SKA \citep[in preparation,][]{Zarka2015}.

\begin{acknowledgements}
PZ acknowledges the support from the Programme National de Plan\'etologie, Programme National Soleil-Terre, and Action Sp\'ecifique SKA-LOFAR of CNRS/INSU, France.  MSM acknowledges the support from CNPq project 154763/2016-0-PDJ, Brazil. EE acknowledges the support from project CNPq/PQ 302583/2015-7, Brazil.  CL acknowledges the support from the Labex Plas@Par, funded by the Agence Nationale de la Recherche as part of the programme "Investissements d'avenir" under the reference ANR-11-IDEX-0004-02. The Nan\c{c}ay Radioastronomy Station acknowledges the support of the Conseil R\'egional of the R\'egion Centre in France. The Nan\c{c}ay Decameter Array acknowledges the support from the Programme National de Plan\'etologie and the Programme National Soleil-Terre of CNRS-INSU.
\end{acknowledgements}


\end{document}